\documentclass[aps,prb,twocolumn,twoside,a4paper,floatfix,showpacs,nobibnotes]{revtex4-1}
\usepackage{graphicx}

\begin{document} 
\input epsf 
\title{Structural Properties of High and Low Density 
Water in a Supercooled Aqueous Solution of Salt} 
\author{
D.~Corradini, M.~Rovere and P.~Gallo}
\email[Author to whom correspondence
should be addressed, e-mail: ]{gallop@fis.uniroma3.it}
\affiliation{Dipartimento di Fisica, Universit\`a ``Roma Tre'', \\ 
Via della Vasca Navale 84, I-00146 Roma, Italy\\}

\begin{abstract}
\noindent 
We consider and compare the structural properties of bulk TIP4P water and of a sodium chloride aqueous solution in TIP4P water with concentration c = 0.67 mol/kg, in the metastable supercooled region. In a previous paper [D. Corradini, M. Rovere and P. Gallo, J. Chem. Phys. {\bf 132}, 134508 (2010)] we found in both systems the presence of a liquid-liquid critical point (LLCP). The LLCP is believed to be the end point of the coexistence line between a high density liquid (HDL) and a low density liquid (LDL) phase of water.
In the present paper we study the different features of water-water structure in HDL and LDL both in bulk water and in the solution. We find that the ions are able to modify the bulk LDL structure, rendering water-water structure more similar to the bulk HDL case. By the study of the hydration structure in HDL and LDL, a possible mechanism for the modification of the bulk LDL structure in the solution is identified in the substitution of the oxygen by the chloride ion in oxygen coordination shells.
\end{abstract}




\maketitle

\section{Introduction}\label{intro}

The possible existence of a second critical point of water in the liquid supercooled metastable phase
has been the subject of a long debate in the literature.
The first hypothesis of its existence originated from the results of a computer simulation 
on water modeled with the ST2 potential~\cite{poole92nature}. On the basis of those results the thermodynamic 
anomalies of water upon supercooling were interpreted in terms of the long range fluctuations induced by the presence of a second critical point. 
This critical point would be a liquid-liquid critical point (LLCP) located at the end of the coexistence 
line between a low density liquid (LDL) phase and a high density liquid (HDL) phase of water. 
In the LLCP scenario, these liquid 
phases would be the counterpart at higher temperature of the well-known low density amorphous (LDA)
and high density amorphous (HDA) phases of glassy water.
The hypothesis of a LLCP scenario for water motivated a large number of experimental, computational and theoretical 
investigations~\cite{pabloreview,deben03}.

Different interpretations of the origin of the thermodynamic anomalies of water have been also proposed as alternatives to the LLCP scenario. In the singularity free scenario~\cite{sastry96} the anomalies of water are due to local density fluctuations and no critical phenomena take place. Recently a critical point free scenario~\cite{angell08} has also been proposed in which the transition between HDL and LDL is seen as an order-disorder transition without a critical point.

A number of computer simulations, performed on supercooled water with different model potentials, confirmed 
the plausibility of the LLCP scenario~\cite{poole93,poole95,harrington97,yamada02,paschek05,paschek08,vallauri05,tanaka96,liu09,sciortino97}.
There are also indications from experiments of the existence of the LLCP in bulk water~\cite{mishima98,banarjee09,mallamace08,mishima10}. 
It would be approximately located at $T \sim 220$~K at $P \sim 100 $~MPa.

Because of the difficulties of performing experiments in the region where the LLCP would reside in bulk water, 
the possibility of observing the LLCP of water in aqueous solutions that can be more easily 
supercooled~\cite{miyata02} has been recently explored
theoretically~\cite{chatterjee} and in computer 
simulations~\cite{corradini10jcp,corradini10pre}.
Results compatible with the existence of a LLCP 
have been also found in aqueous solutions of salts through thermometric 
experiments~\cite{archer00,archer00pccp,mishima05,mishima07}. 

In a recent paper~\cite{corradini10jcp}, by means of a 
computer simulation study of the phase diagram, we indicated the possible detection in 
thermometric experiments of the LLCP in a NaCl(aq) solution. 

Since the detection of low and high density forms of water 
can also offer a viable path to the experimental detection of a LLCP,
structural properties of supercooled water and aqueous solutions are of extreme interest in this context.
The structure of water and of aqueous solutions, can be studied with neutron diffraction
using isotopic substitution~\cite{soper00,mancinelli07jpcb} or by X-ray scattering~\cite{neilson01}. 

In the present paper we focus on the structural properties of bulk TIP4P water and of 
the NaCl(aq) solution with $c=0.67$~mol/kg, in order to analyze and compare the results in HDL and LDL especially close to the LLCP.
The paper is organized as follows. In Sec.~\ref{details} the details of the computer simulations
are given. In Sec.~\ref{thermo} we summarize the main results obtained on
the thermodynamics of bulk water and NaCl(aq) and we present the potential energy
of the systems. The new results for the structural properties of the systems are presented in Sec.~\ref{results}.
This section is divided in two parts: water-water structure is discussed in subsection~\ref{water}, 
while the hydration 
structure of ions is addressed in subsection~\ref{hydration}. 
Finally, conclusions are given in Sec.~\ref{conclusions}.

\section{Computer simulation details}\label{details}

Molecular dynamics (MD) computer simulations were performed on bulk water and on 
NaCl(aq) with concentration $c=0.67$~mol/kg. The interaction potential between pairs 
of particles is given by the sum of the electrostatic and the Lennard-Jones (LJ) potentials.
\begin{equation}\label{eq:1}
U_{ij}(r)=\frac{q_i q_j}{r}+4\epsilon_{ij}\left[ \left(\frac{\sigma_{ij}}{r}\right)^{12}- \left(\frac{\sigma_{ij}}{r}\right)^6\right]
\end{equation}
Water molecules were modeled using the TIP4P potential~\cite{jorgensen83}. The details about this 
potential are reported in the Appendix. 
TIP4P potential is known to well describe
water properties in its liquid 
supercooled state~\cite{jorgensen83} and also 
to be able to reproduce the very complex ices phase diagram~\cite{sanz04}.

The LJ interaction parameters 
for the ions were taken from Jensen and Jorgensen~\cite{jensen06} and the ion-water interaction parameters
were calculated by using geometrical mixing rules
$\epsilon_{ij}=(\epsilon_{ii} \epsilon_{jj})^{1/2}$ and $\sigma_{ij}=(\sigma_{ii} \sigma_{jj})^{1/2}$.
The ion-ion and ion-water parameters are reported in Table~\ref{tab:1}. These parameters were optimized for use with
TIP4P water and they well reproduce structural characteristics and free energies of hydration of the ions~\cite{jensen06}.

Although the presence of ions would suggest the use of
polarizable potentials, at the moment no joint set
of polarizable potentials for water and ions are tested 
as reliable for very low temperatures.

\begin{table}[htdb]
\caption{Ion-ion and ion-water LJ interaction parameters~\cite{jensen06}.}
\begin{center}
\begin{ruledtabular}
\begin{tabular}{lcc} Atom pair & $\epsilon\, (\mbox{kJ/mol})$ &
$\sigma (\mbox{\AA})$\\ 
\hline 
Na-Na&0.002&4.070\\
Na-Cl&0.079&4.045\\
Cl-Cl&2.971&4.020\\
Na-O&0.037&3.583\\
Cl-O&1.388&3.561\\
\end{tabular}
\end{ruledtabular}
\end{center}
\label{tab:1}
\end{table}

Periodic boundary conditions were applied. The cutoff radius was set at 9~\AA. Standard 
long-range corrections were applied for the calculation of the potential energy and the virial. 
The long-range electrostatic interactions were handled with the Ewald summation method. The integration 
time step was fixed at 1~fs.

The total number of particles contained in the simulation box is $N_{tot}=256$. For bulk water 
$N_{wat}=N_{tot}=256$ while in the case of NaCl(aq) with concentration $c=0.67$~mol/kg, 
$N_{wat}=250$ and $N_{Na^+}=N_{Cl^-}=3$. Extensive sets of simulations were run both for bulk 
water and for NaCl(aq). The range of densities investigated spans from $\rho=0.83\, g/cm^3$ to 
$\rho=1.10\, g/cm^3$ and the range of temperatures goes from $T=350$~K to $T=190$~K. 
Temperature was controlled using the Berendsen thermostat~\cite{berendsen}.
Equilibration and production simulation times were progressively increased with the decreasing temperature.
The total running times span from 0.15 ns for the highest temperatures to 30 ns for the lowest ones.
The parallelized version of the DL\_POLY package~\cite{dlpoly} was employed to perform the simulations. 
The total simulation time is circa 6 single CPU years.

\section{Thermodynamics in the supercooled region}\label{thermo}

In this section we start by summarizing 
the main results on the thermodynamic behavior in the supercooled 
region of bulk water and of NaCl(aq) obtained in Ref.~\onlinecite{corradini10jcp} and then we analyze
the behavior of the potential energy of the systems 
studied at different temperatures.
From extensive simulations on bulk TIP4P water and on NaCl(aq) 
with concentration $c=0.67$~mol/kg, we located in both systems the LLCP. 
Details about the calculation of the position of the LLCP are given in Ref.~\onlinecite{corradini10jcp}. 

\begin{figure}[!t]
\includegraphics[width=\columnwidth,clip]{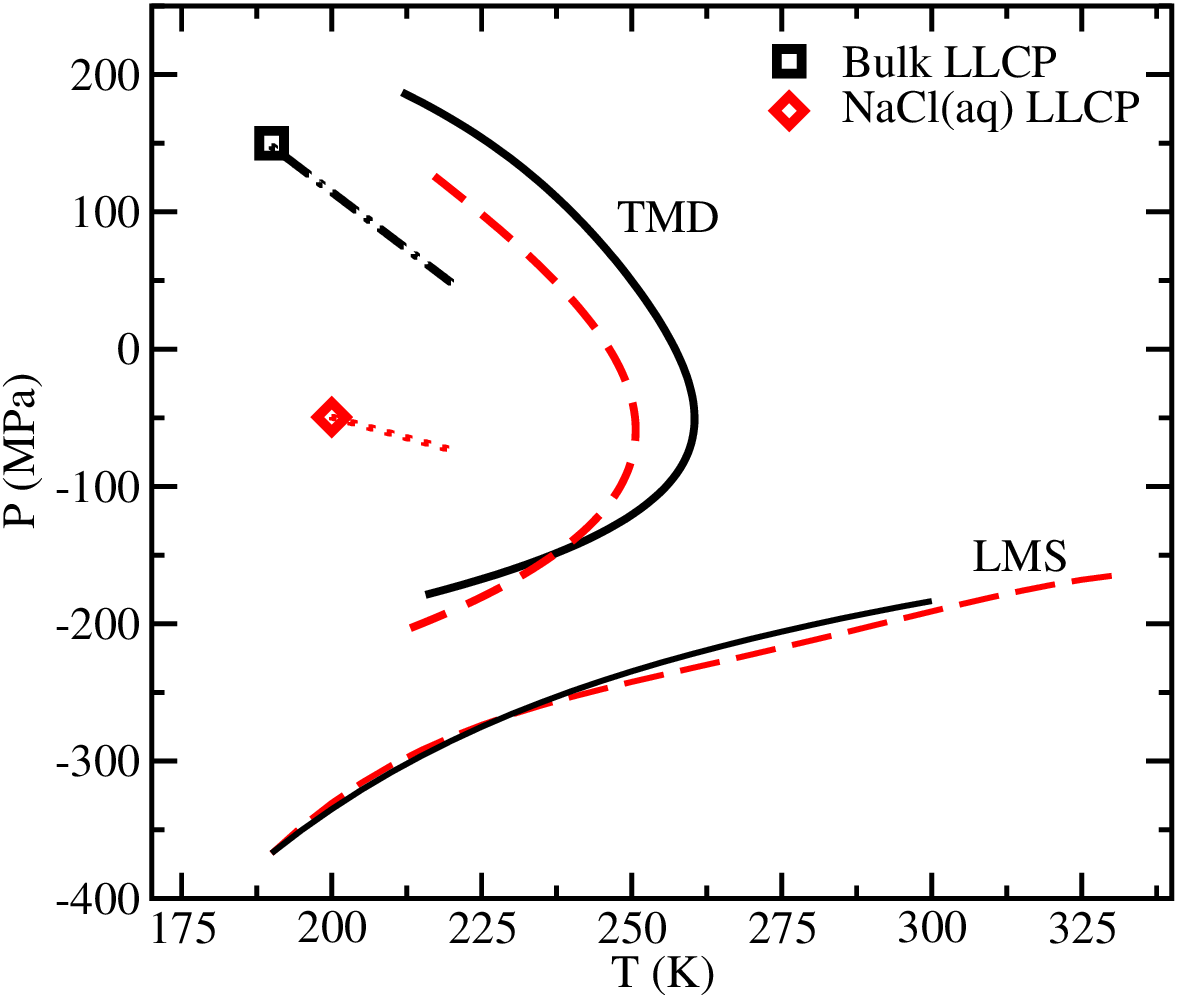}
\caption{Comparison  between the thermodynamic features of bulk water and NaCl(aq) in
the supercooled region, based on the results shown in Ref.~\onlinecite{corradini10jcp}.
We report the position of the LLCP and of the Widom line for bulk water (dot-dashed line) and for NaCl(aq)
(dotted line). The TMD and LG-LMS lines are also shown for bulk water (solid lines) and NaCl(aq) (dashed lines).}
\label{fig:1}
\end{figure}

In Fig.~\ref{fig:1} 
we report a comparison of the phase diagrams of the supercooled regions
of bulk water and NaCl(aq) as obtained directly from MD. Together with the LLCP of the two systems, we show
the Widom lines, the temperature of maximum density (TMD) lines and the liquid-gas
limit of mechanical stability (LG-LMS) lines. The Widom line can be considered an
extension of the coexistence line in the one-phase region~\cite{franzese07,xu05}.
In bulk water the position of the LLCP is $T_c=190$~K and $P_c=150$~MPa.
In a very recent paper~\cite{abascal10} another estimate of the LLCP for bulk water has been performed
with a TIP4P potential with modified parameters~\cite{abascal05}. 
The values that the authors obtained for the critical point,
$T_c=193$~K and $P_c=135$~MPa, and the Widom line 
appears substantially the same as the values that we found 
with the original TIP4P~\cite{corradini10jcp}. 

Coming back to Fig.~\ref{fig:1} 
we can see that in NaCl(aq)  the position of the
LLCP moves to lower pressure and higher temperature, appearing at $T_c=200$~K and $P_c=-50$~MPa.
The TMD line of the solution lies circa 10~K
below in temperature and at slightly lower pressure with respect to bulk water. The LG-LMS instead
is almost unchanged with respect to the bulk. From the comparison of the phase
diagrams for these two systems, we found that the main effect of the presence 
of the ions is to shrink the region of existence of the LDL~\cite{corradini10jcp}, 
consistent with an observed increased solubility of ions in HDL water~\cite{mishima07,souda}.

Importantly, we also found that with a rigid shift in temperature and pressure of the phase diagram 
of TIP4P bulk water (not shown), that brought the TMD curve to coincide with the experimental TMD values,
the LLCP in bulk water appears at $T_c=221$~K and $P_c=77$~MPa, close to the 
experimental value estimated by Mishima and Stanley~\cite{mishima98} 
$T_c\sim 220$~K and $P_c\sim 100$~MPa.  We note that this shift in temperature is compatible with the 
shift between the melting line of TIP4P~\cite{vega05} and 
the melting line of real water.
Recently Mishima published a new estimate of the LLCP in the bulk at $T_c\sim 223$~K and $P_c\sim 50$~MPa~\cite{mishima10}
also compatible with our findings. 
These results confirm that TIP4P is a good potential 
for describing the supercooled water phase diagram.
The same shift applied to our ionic solution led us to predict a 
LLCP in NaCl(aq) located at $T_c\sim 231$~K and $P_c\sim -123$~MPa. 
These last values appear to be in a region accessible by experiments, 
being above the homogeneous nucleation temperature of the solution~\cite{miyata02,kanno75}.

\begin{figure}[!t]
\includegraphics[width=\columnwidth,clip]{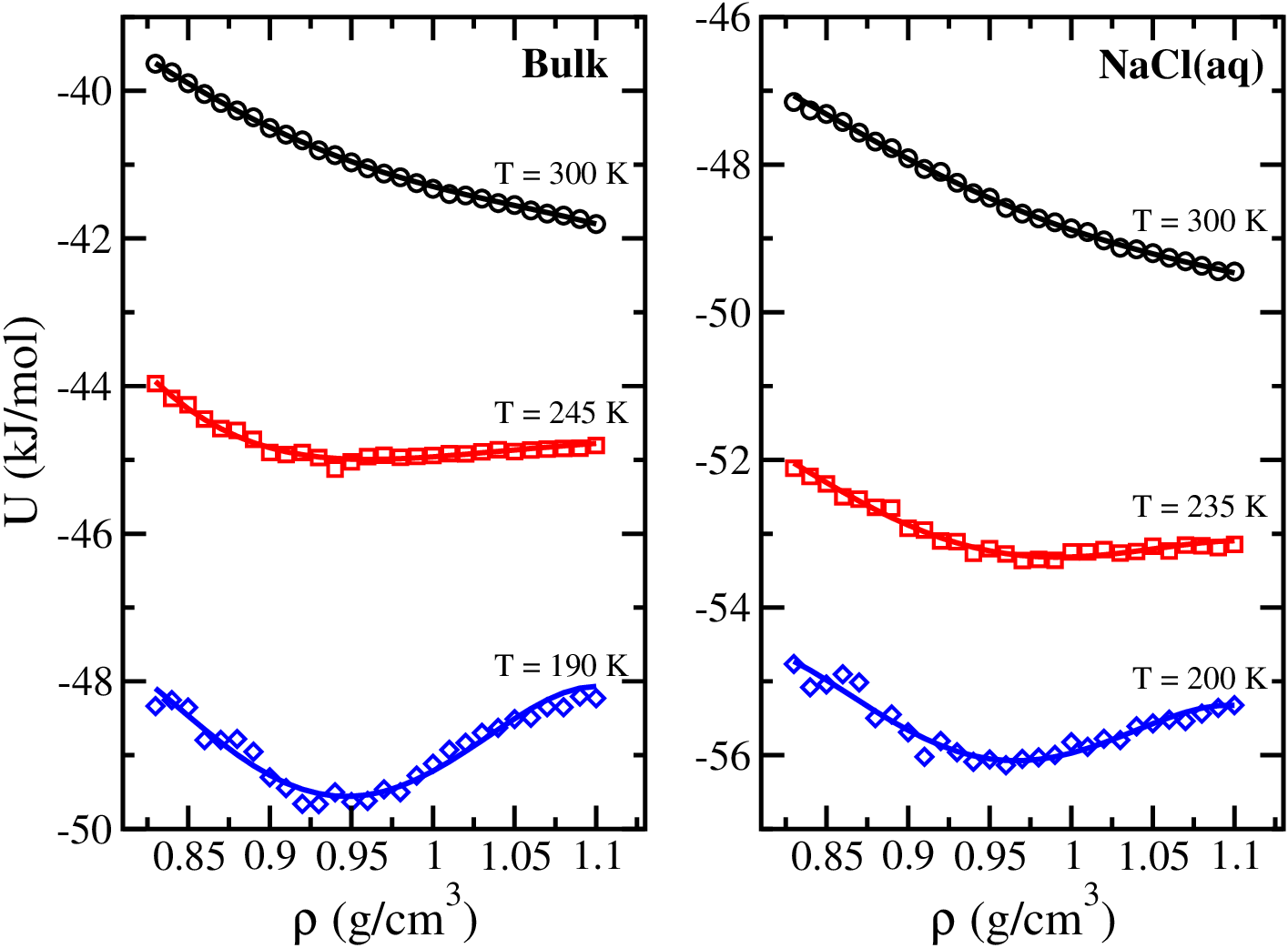}
\caption{Potential energy per molecule as a function of the density of the system at constant
temperature. For bulk water (left panel) the isotherms of the potential energy are reported at
$T=300$~K, $T=245$~K and $T=190$~K. For NaCl(aq) (right panel) the isotherms of the 
potential energy are reported at $T=300$~K, $T=235$~K and $T=200$~K.
Continuous lines are a polynomial best fits to the simulated points.}
\label{fig:2}
\end{figure}

Generally speaking, from the behavior of the potential 
energy $U$ it is also possible to extract information 
on the thermodynamics of a system close to a phase transition.
As already pointed out in 
Ref.~\onlinecite{harrington97} and~\onlinecite{sciortino97} 
if we examine the curvature 
of the configurational part of the Helmholtz free energy $A=U-TS$,
\begin{equation}
\left(  \frac{\partial^2 A}{\partial V^2}\right)_T 
=\left( \frac{\partial^2 U}
{\partial V^2} \right)_T - T \left( 
\frac{\partial^2 S}{\partial V^2} \right)_T  
\label{eq:2}
\end{equation}
it must be positive in the region of stability of an homogeneous phase. 
When the curvature of $U$ is negative, the system can still be stable due to the contribution of a dominant entropic term in Eq.~(\ref{eq:2}).
At supercooled low temperatures, the stabilization induced by the entropic term
 can be less effective as the factor $T$ in front of the 
second derivative of the entropy in Eq.~\ref{eq:2} becomes 
progressively smaller.
Thus, at low temperatures
the range of volumes where $(\partial^2 U/\partial V^2)_T<0$,
corresponds to a region of reduced stability for the homogeneous liquid,
where separation in two distinct phases with different densities may occur.

In Fig.~\ref{fig:2} the isotherms of the potential energy are reported for three
temperatures for bulk water and NaCl(aq). We show the behavior
of potential energy for ambient temperature $T=300$~K,  for the first
temperature showing a negative curvature of $U$, $T=245$~K in the 
bulk and $T=235$~K in the solution, and for the LLCP temperature,
$T=190$~K in the bulk and $T=200$~K in the solution.
In both systems, at high temperature the potential energy is
a positively curved function of the density and it decreases
monotonically (becomes more negative) as density increases.
When the temperature is decreased, a minimum is formed after which the curvature
of the potential energy becomes negative. The minimum corresponds to
the presence of a tetrahedrally ordered liquid with an energetically favorable
configuration~\cite{sciortino97}. For the temperature corresponding to
that of the LLCP, we see that in both systems the minimum
becomes very deep and a maximum is suggested at high densities, indicating
the possible occurrence of a second minimum at higher densities.
This second minimum has been previously observed for confined bulk water~\cite{kumar05}
and for higher concentration NaCl(aq) solutions~\cite{corradini09} and connected
to the existence of two distinct liquid phases in the system, as at very low temperatures
the entropic contribution is depressed and the behavior of the free energy $A$ can be
approximated with that of the potential energy $U$.
Comparing the behavior of the potential energy of bulk water and NaCl(aq), we can
notice that apart from the shift in the absolute value due the presence of ions,
their behavior is quite similar. Nonetheless the minimum
at low density becomes more shallow in the solution, indicating that this phase is made
less stable by the presence of ions, consistent with the fact that ions stabilize the 
high density phase~\cite{corradini10jcp}.
 
\section{Structural results}\label{results}

As discussed in the previous section, the study of the phase diagram of bulk water and of the aqueous solution shows the
presence of a LLCP in the supercooled region. 
We now discuss the structural properties of the systems. In the following subsection, we analyze water-water structure both in the bulk and in the solution. Then we study the hydration (ion-water) structure.

\subsection{Water-water structure}\label{water}

\begin{figure}[!t]
\includegraphics[width=\columnwidth,clip]{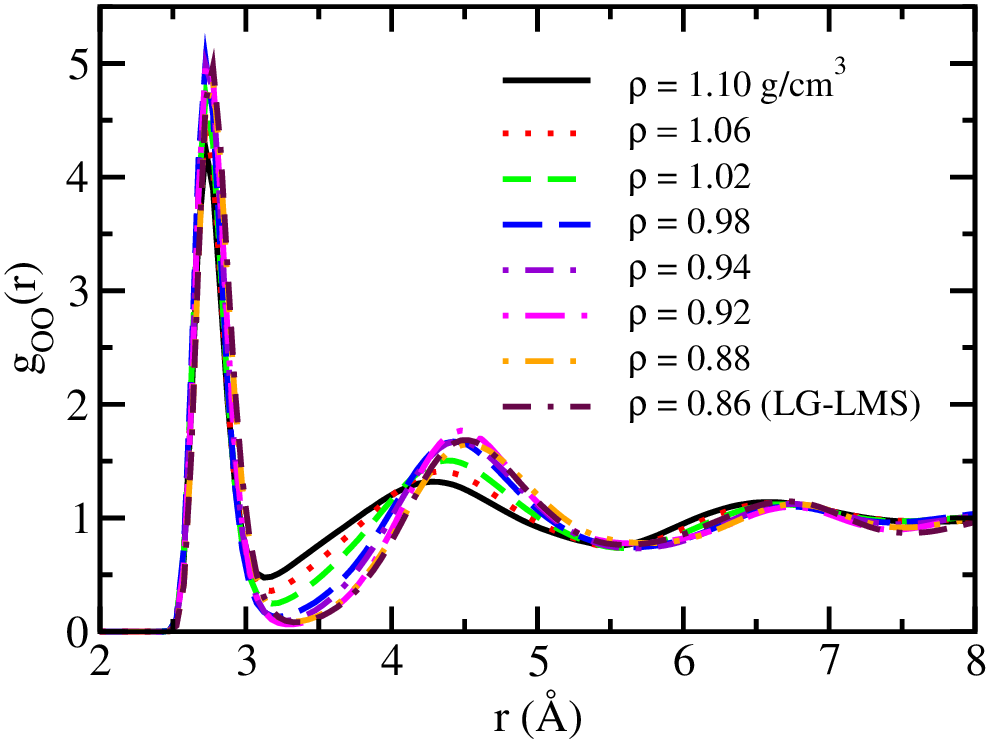}
\caption{O-O RDFs of bulk water at $T=190$~K
for densities from $\rho=1.10\, g/cm^3$
to $\rho=0.86\, g/cm^3$ (LG-LMS).}
\label{fig:3}
\end{figure}

In Fig.~\ref{fig:3} we report the O-O radial distribution functions (RDFs) of bulk water at the LLCP 
temperature for decreasing densities from $\rho=1.10\, g/cm^3$ to the LG-LMS density $\rho=0.86\, g/cm^3$.
According to our phase diagram, the LLCP in bulk water is located at 
$\rho=1.06\, g/cm^3$.
As a general trend, there is an increase of the first peak with decreasing density. 
We note that the second peak 
from $\rho=0.86\, g/cm^3$ to $\rho=0.98\, g/cm^3$ does not substantially change position. 
As the LLCP is approached, from $\rho=1.02\, g/cm^3$ it starts to shift to lower distances. 

The LLCP is located at $\rho=0.99\, g/cm^3$ in NaCl(aq);
thus in order to study the structural differences between the LDL and the HDL both in bulk water and in NaCl(aq),
we will take into account, in the following discussion, the RDFs at two density values which are well above
and well below the estimated critical densities, namely $\rho=1.10\, g/cm^3$ for HDL and
$\rho=0.92\, g/cm^3$ for LDL.

\begin{figure}[!t]
\includegraphics[width=\columnwidth,clip]{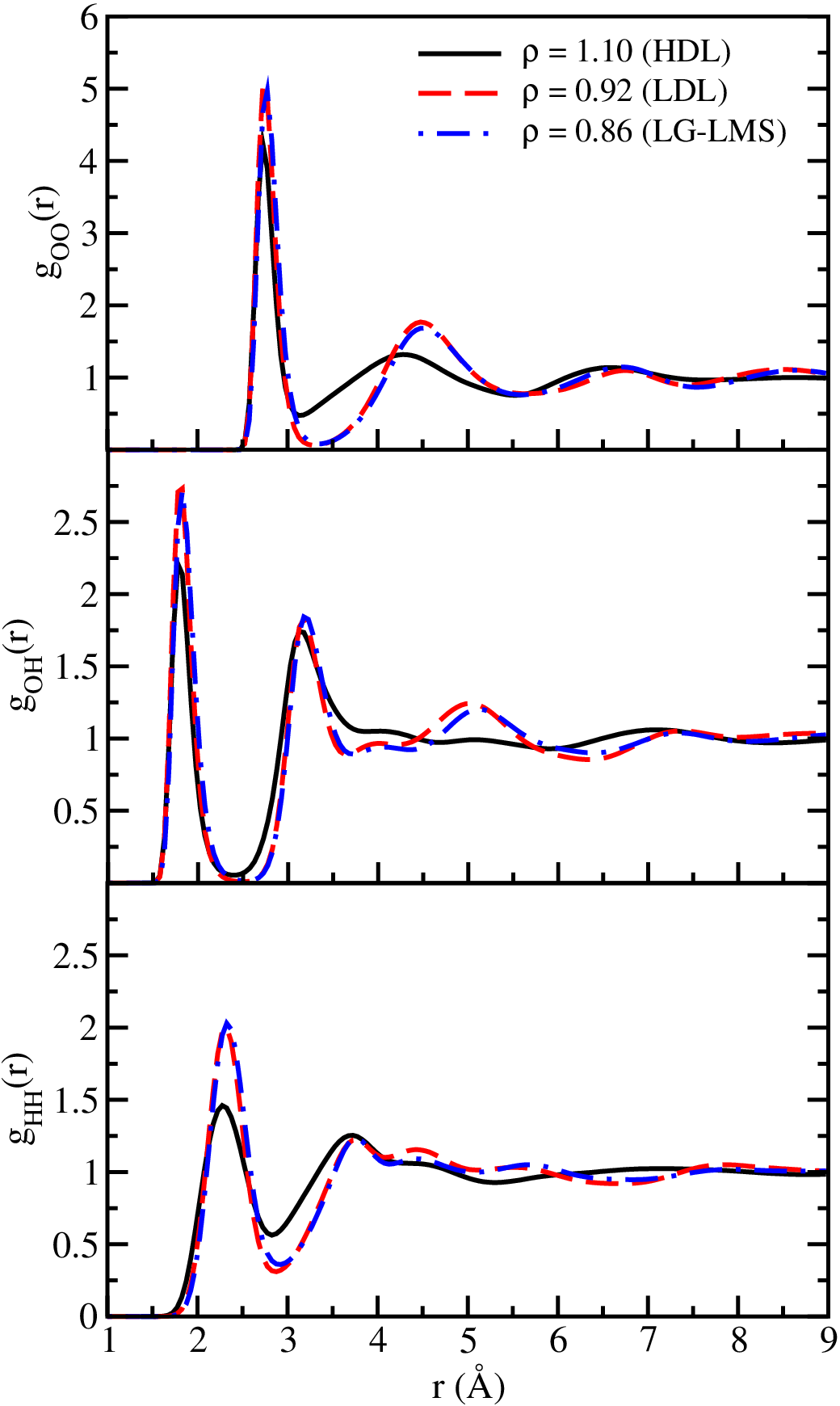}
\caption{O-O (top panel), O-H (central panel) and H-H (bottom panel) RDFs for bulk water at $T=190$~K. 
Solid lines: $\rho=1.10\, g/cm^3$ (HDL); dashed lines : $\rho=0.92\, g/cm^3$ (LDL);
dot-dashed lines: $\rho=0.86\, g/cm^3$ (LG-LMS). }
\label{fig:4}
\end{figure}

In Fig.~\ref{fig:4} we compare the water-water RDFs $g_{OO}(r)$, $g_{OH}(r)$ and $g_{HH}(r)$ 
of bulk water obtained at the thermodynamic conditions in which water is either in the LDL or in the HDL region. 
The RDFs are plotted for $T=190$~K at densities $\rho=0.92\, g/cm^3$ and $\rho=1.10\, g/cm^3$ representative of LDL and HDL respectively. 
The RDFs at the density $\rho=0.86\, g/cm^3$ that represents, at this
temperature, the LG-LMS of bulk water are also reported for comparison.
We observe for the O-O RDF that the height of the first peak decreases going from 
LDL to HDL while its position does not change.
The second peak instead is markedly different. In HDL its position shifts to lower distances,
its height decreases and its shape broadens, with respect to the LDL.
For the $g_{OH}(r)$ the height of the first peak decreases in HDL and it moves to
slightly lower distances. This slight shift is conserved between the first and the second shell,
while the position and the height of the second shell are fairly similar for HDL and LDL, 
apart from the appearance of a shoulder between the second and the third shell in HDL.
It can be also noted that while the LDL shows a very well-defined third shell, it disappears in
HDL. For the $g_{HH}$(r) also the first peak of the HDL is less intense than the one of LDL and
slightly shifted to lower distances. It can also be noted a widening of the second shell
in the HDL.
The overall trend of water-water RDFs and in particular the difference in the second shell
of the O-O RDF clearly shows the disruption of hydrogen bonds between the first and
second shell of water molecules that occurs in HDL, causing it to have a collapsed
second shell with respect to the tetrahedrally coordinated LDL~\cite{soper00}.

\begin{figure}[!t]
\includegraphics[width=\columnwidth,clip]{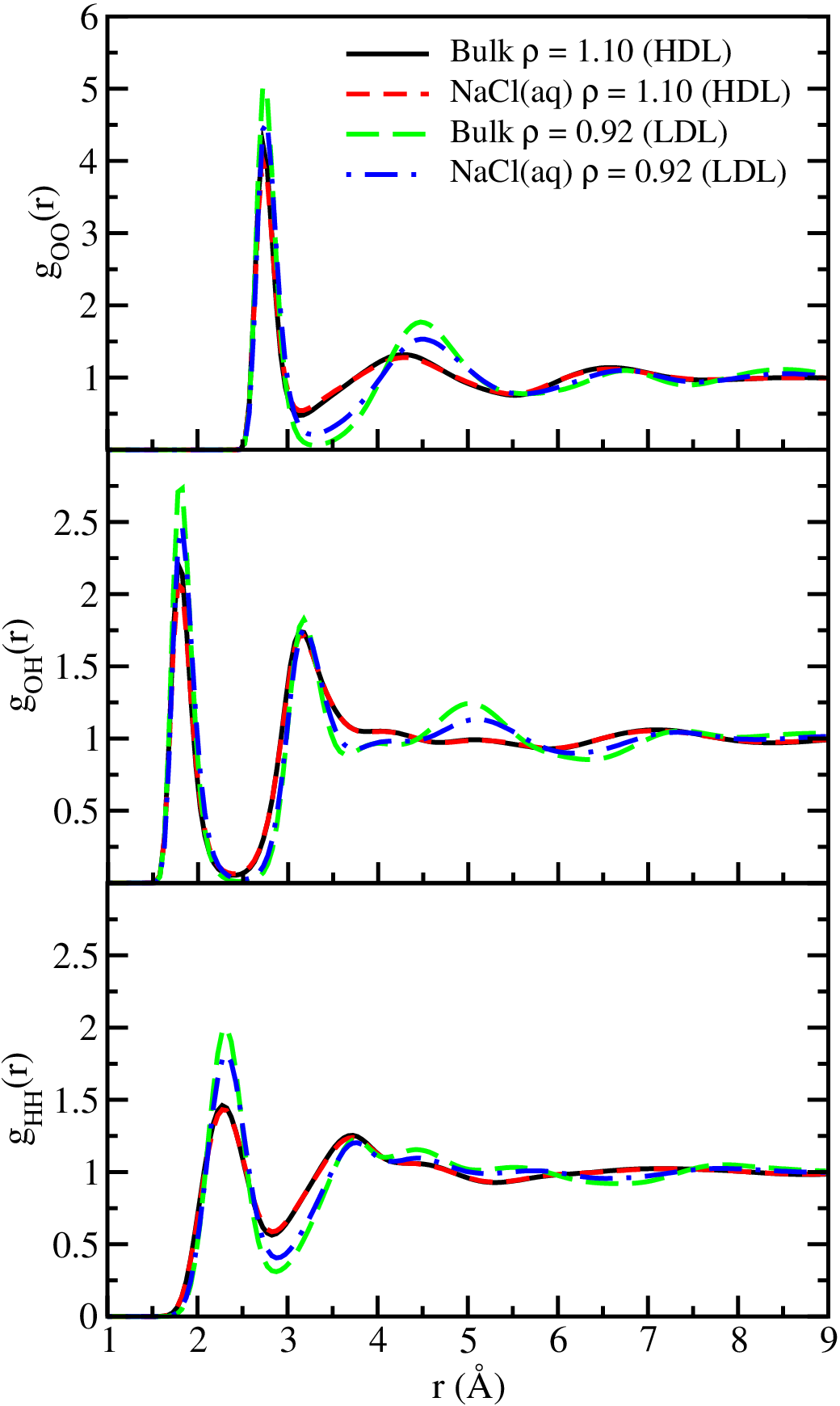}
\caption{O-O (top panel), O-H (central panel) and H-H (bottom panel) RDFs for bulk water at $T=190$~K 
and for NaCl(aq) at $T=200$~K.
Solid lines: bulk $\rho=1.10\, g/cm^3$ (HDL); dashed lines:  NaCl(aq) $\rho=1.10\, g/cm^3$ (HDL);
long dashed lines: bulk $\rho=0.92\, g/cm^3$ (LDL); dot-dashed lines: NaCl(aq) $\rho=0.92\, g/cm^3$ (LDL).}
\label{fig:5}
\end{figure}

The results for HDL and LDL RDFs in bulk water are in good agreement
with those found in experiments~\cite{soper00}. 
It has been shown that the TIP4P model is a very reliable model
for the study of the thermodynamics of solid~\cite{sanz04} and supercooled liquid water~\cite{corradini10jcp}.
Our results confirm the validity of the TIP4P model also for the study of the structural properties in the
supercooled region.

In Fig.~\ref{fig:5} we compare the HDL and LDL water-water RDFs for bulk water
and for NaCl(aq). In NaCl(aq) the overall shape and positions of the peaks remain similar
to those of bulk water in both cases, but some differences can be noted for the LDL. 
The height of the first peak of $g_{OO}(r)$, $g_{OH}(r)$ and $g_{HH}(r)$ is 
slightly lower for the NaCl(aq) LDL with respect to the correspondent phase
in bulk water. Furthermore in the $g_{OO}(r)$ the first minimum moves to lower distances in the solution
and the height of the second peak is reduced. In the $g_{OH}(r)$ the first and the second 
shell remain similar but the height of the third shell is damped in the NaCl(aq).
These results indicate that at this concentration the effect of ions on water-water structure
is not strong and all the features found in bulk water are preserved. Nonetheless,
LDL seems to be more affected by the presence of ions than
HDL. This can be due to a certain degree of disruption of hydrogen bonds induced by the
ions. In fact, as shown in Fig.~\ref{fig:5} we see that the effect of ions is that of reducing
the LDL character of water-water structure and making it more similar to the one
of the HDL. These results are in agreement with that observed in our study the thermodynamics
of these systems~\cite{corradini10jcp}. In fact, it was found that the range of existence of the 
LDL phase is reduced when ions are added to bulk water.

\begin{figure}[!t]
\includegraphics[width=\columnwidth,clip]{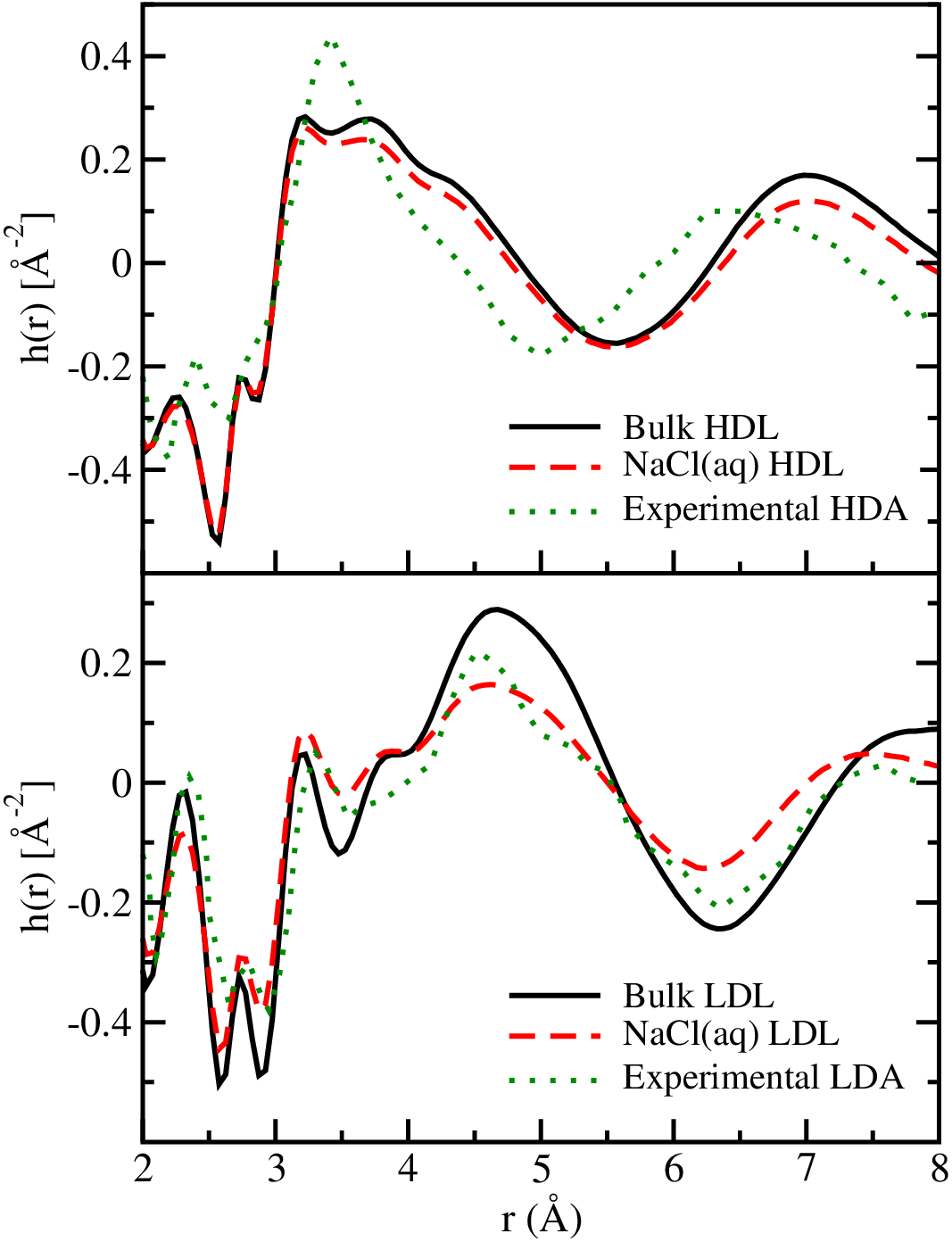}
\caption{Function $h(r)$, see text for definition. Top panel:
bulk HDL (solid line), NaCl(aq) HDL (dashed line) and
experimental HDA (dotted line). Bottom panel:
bulk LDL (solid line), NaCl(aq) LDL (dashed line) and experimental LDA
(dotted line).}
\label{fig:6}
\end{figure}

To complete the comparison between HDL and LDL in bulk water and in NaCl(aq)
we report in Fig.~\ref{fig:6} the quantity
\begin{equation}
h(r)=4 \pi \rho r [ 0.092 g_{OO}(r) + 0.422 g_{OH}(r)
+ 0.486 g_{HH}(r) -1 ]
\end{equation}
This correlation function has been obtained in neutron
diffraction experiments on the amorphous HDA and LDA phases 
of water~\cite{bellissent92,bellissent89}. 
The qualitative agreement shows the existing relation between the
LDA/HDA phases of ice with the corresponding LDL/HDL phases of water 
as already pointed out in simulations of bulk water with the ST2 potential~\cite{sciortino97}.
We observe again that LDL is more influenced by the presence of ions.
The modifications to $h(r)$ in fact appear larger in LDL with a significant reduction of 
the peak at about 4.6~\AA~and the different behavior for long distances, $r>6.5$~\AA. This confirms
the behavior that we observed for LDL, looking at the partial water-water RDFs.

\begin{figure}[!t]
\includegraphics[width=\columnwidth,clip]{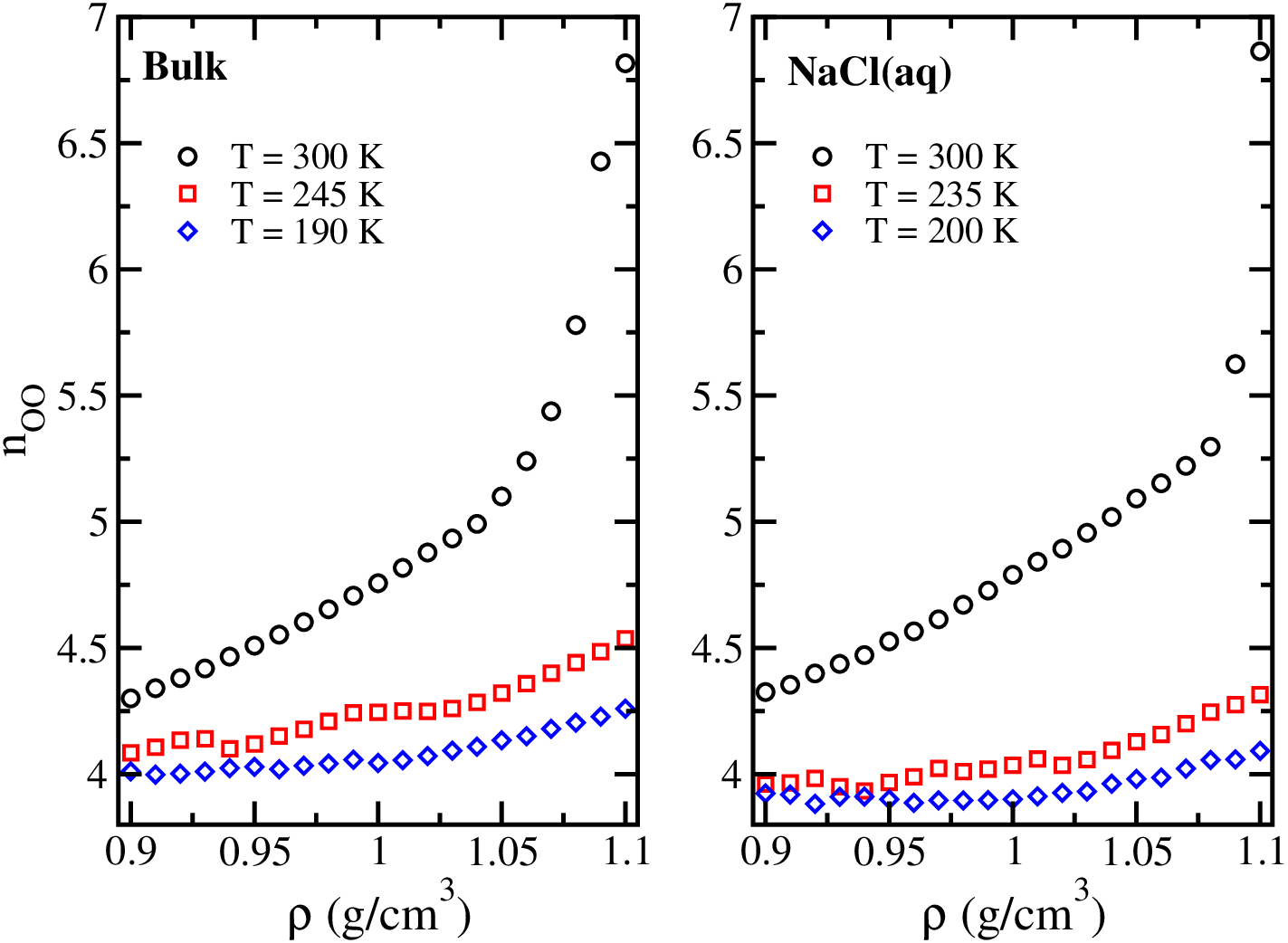}
\caption{O-O first shell coordination number as a function of density, at constant temperature, 
for bulk water (left panel) and NaCl(aq) (right panel). The temperatures reported are $T=300$~K,
$T=245$~K and $T=190$~K for bulk water and $T=300$~K,
$T=235$~K and $T=200$~K for NaCl(aq).}
\label{fig:7}
\end{figure}

To conclude the analysis of water-water structure we report in Fig.~\ref{fig:7}
the oxygen-oxygen first shell coordination numbers as a function of 
density, at constant temperature. We report them at the same temperatures that
we show in Fig.~\ref{fig:2} for the potential energy,  $T=300$~K,
$T=245$~K and $T=190$~K for bulk water and $T=300$~K,
$T=235$~K and $T=200$~K for NaCl(aq). The first shell coordination
numbers can be calculated integrating the radial distribution function
until the first minimum is reached, and thus they represent the average
number of first neighbors around an oxygen atom. At high temperatures
the coordination numbers decrease monotonically with the density,
without reaching the value of 4 typical of LDL~\cite{sciortino97} in the
spanned range of simulated state points. As temperature is decreased,
the isotherms of the coordination numbers approach an asymptotic value
of 4 when density decreases. For the lowest temperature, the coordination number
reaches the value 4 already at quite high densities in bulk water 
and it actually drops slightly 
below 4 in NaCl(aq). 
In the case of the first shell coordination numbers, HDL and LDL
seem to be affected in a similar way by the presence of ions, with a slight
decrease of the average number of first neighbors. Nonetheless, the
weakening of the ordered tetrahedral configuration of LDL in the first shell induced by the ions
may allow for a reorganization of the bonds that leads to a more packed, HDL-like structure,
as we have seen looking at the RDFs.

\subsection{Hydration structure}\label{hydration}

The hydration structure in NaCl(aq) at ambient temperature and in the moderately supercooled region 
has been the subject of several experimental~\cite{mancinelli07jpcb,mancinelli07pccp,ohtaki01} and computer simulation 
works~\cite{corradini08,koneshan00,chowdhuri01-03,patra04,lenart07,alejandre07,lynden01,zhu92,kim08,du07,joung08}. 
Nonetheless, the hydration behavior close to the LLCP has never been studied previously.

\begin{figure}[!t]
\includegraphics[width=\columnwidth,clip]{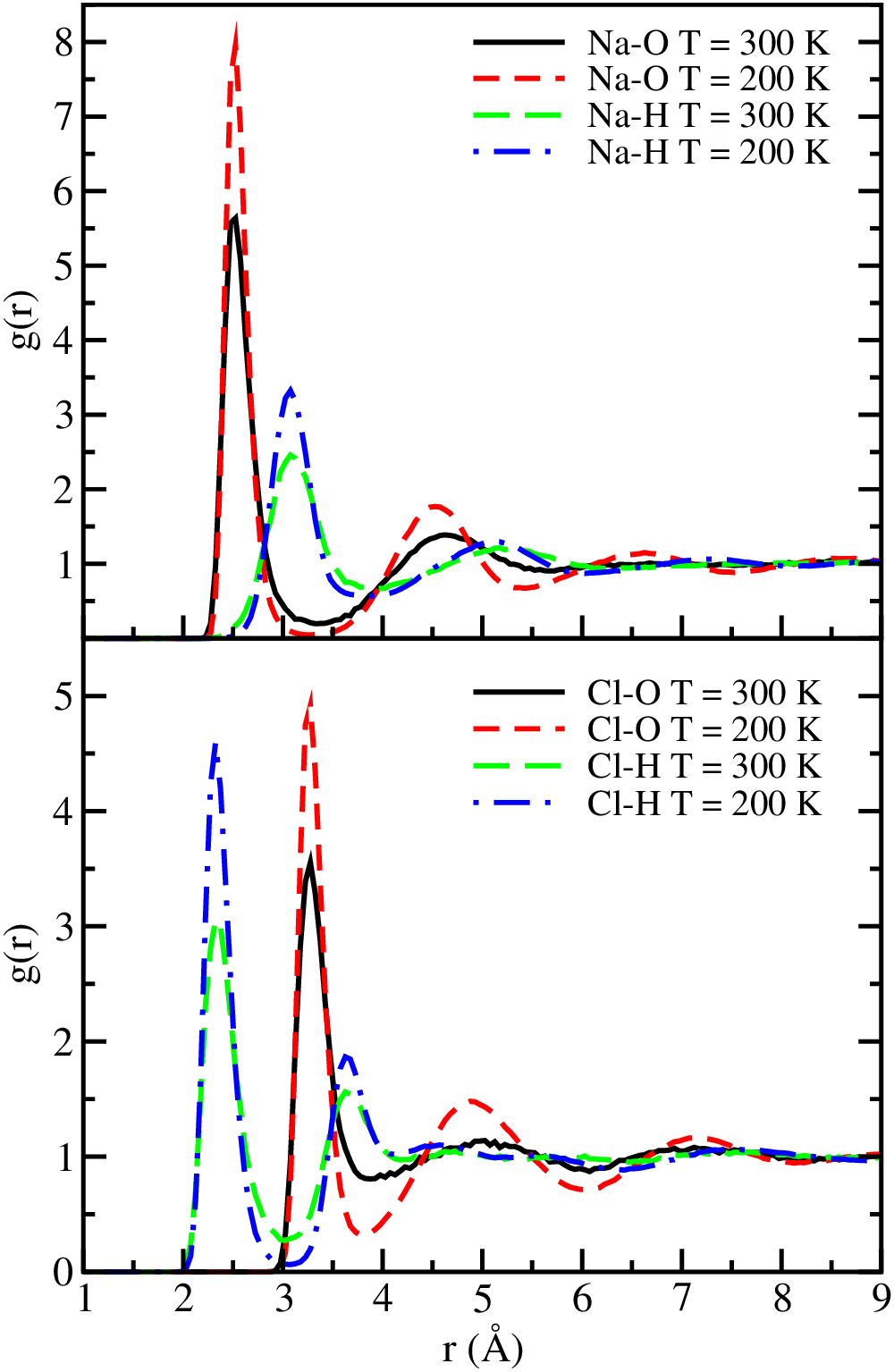}
\caption{Na-water (top panel) and Cl-water (bottom panel) 
RDFs at $\rho=1.00\, g/cm^3$ and at temperatures $T=300$~K and $T=200$~K.}
\label{fig:8}
\end{figure}

\begin{figure}[!t]
\includegraphics[width=\columnwidth,clip]{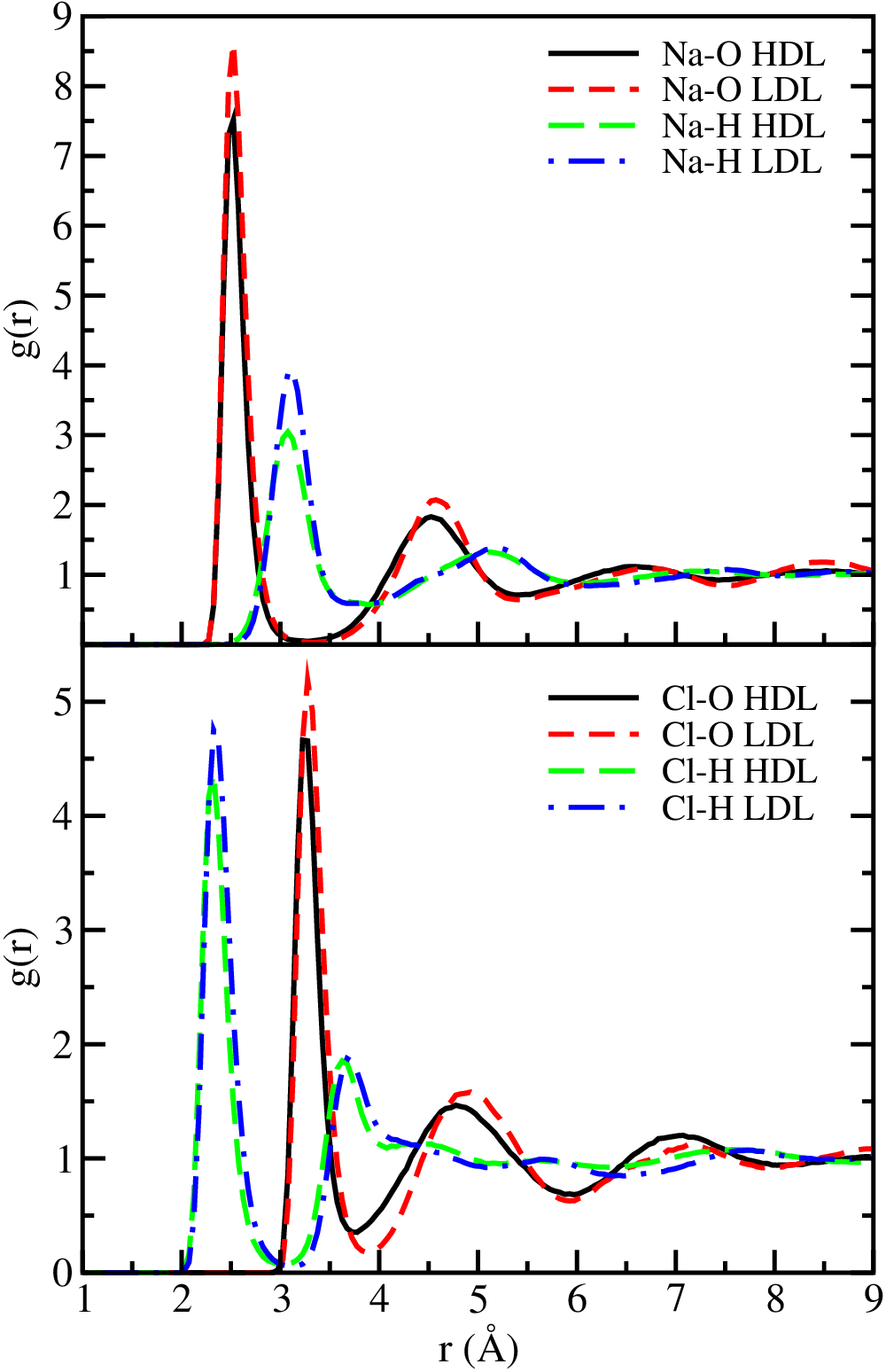}
\caption{Na-water (top panel) and Cl-water (bottom panel) RDFs for $\rho=1.10 \, g/cm^3$ (HDL) and $\rho=0.92 \, g/cm^3$ (LDL), at $T=200$~K.}
\label{fig:9}
\end{figure}

In order to study the general features of the hydration shells we
show in Fig.~\ref{fig:8} the Na-water and Cl-water RDFs for $\rho=1.00\, g/cm^3$
at ambient temperature $T=300$~K and in the deep supercooled region, at
$T=200$~K, corresponding to the LLCP temperature in the solution.
For all the four couples we can observe that there are two well-defined
hydration shells. The height of both the first and second peak
tends to increase greatly upon supercooling. In the case of Na-O and 
Cl-O we observe also for the second peak also a shift of its position
to slightly lower distances at the lowest temperature. These results
seem to indicate that upon supercooling the hydration shells tend to stabilize
and to become more compact.
In the case of the chloride ion, the Cl-O first and second peaks are very well separated from the
respective Cl-H peaks, by roughly 1~\AA. In the case of the sodium ion the distance is reduced to roughly
0.6~\AA~and there is actually an overlap of the Na-O and Na-H hydration
shells.
For the chloride ion the hydration structure of the shells is similar
to the case of O-O and O-H structure (see Fig.~\ref{fig:4} and Fig.~\ref{fig:5}).
Thus, we argue that chloride can substitute oxygen in the hydration shells
forming linear hydrogen bonds with oxygen atoms as indicated by the sharp, 
first Cl-H peak.
The Na$^+$ ion cannot simply substitute the oxygen atom. Its positive charge induces
a major rearrangement of the hydration shells that results in more packed structure.
The behavior of the hydration shells of Na$^+$ and Cl$^-$ ions is in good agreement
with those found experimentally for ambient temperature in recent
neutron diffraction experiments~\cite{mancinelli07jpcb,mancinelli07pccp}.

In Fig.~\ref{fig:9} we report the Na-water and Cl-water RDFs for the HDL, at
$\rho=1.10\, g/cm^3$ and the LDL at $\rho=0.92\, g/cm^3$ and at $T=200$~K.
We can see that the hydration shells of sodium ions are not much affected
by the HDL or LDL environment of the solvent. An increase of the first
and second peak of the Na-O RDF and of the first peak of the Na-H RDF
is observed in LDL with respect to HDL but the position of the peaks
remains unaltered apart from a very slight shift to higher distances of 
the second shell of Na-O for the LDL.
In the case of the chloride ion, the Cl-H RDF remains unchanged in
HDL or LDL but the Cl-O RDF show some differences. In the LDL the first peak is higher
than in HDL and the second peak is both higher and shifted to longer distances.
We report in Table~\ref{tab:2} the first shell and second shell
hydration numbers for the sodium and chloride ions at the LLCP temperature.
These numbers have been calculated by numerical integration of the first and 
second peak of the RDFs plotted in Fig.~\ref{fig:9}. The integration range 
spans form zero to the first minimum for the first hydration shell and from 
the first to the second minimum for the second hydration shell. 
The modifications of the first shell in going from HDL to LDL
are practically imperceptible. The second shell appears already sensitive
to the HDL and LDL environment as we can deduce from the decrease of hydration 
numbers in going from HDL to LDL.

\begin{figure}[!t]
\includegraphics[width=\columnwidth,clip]{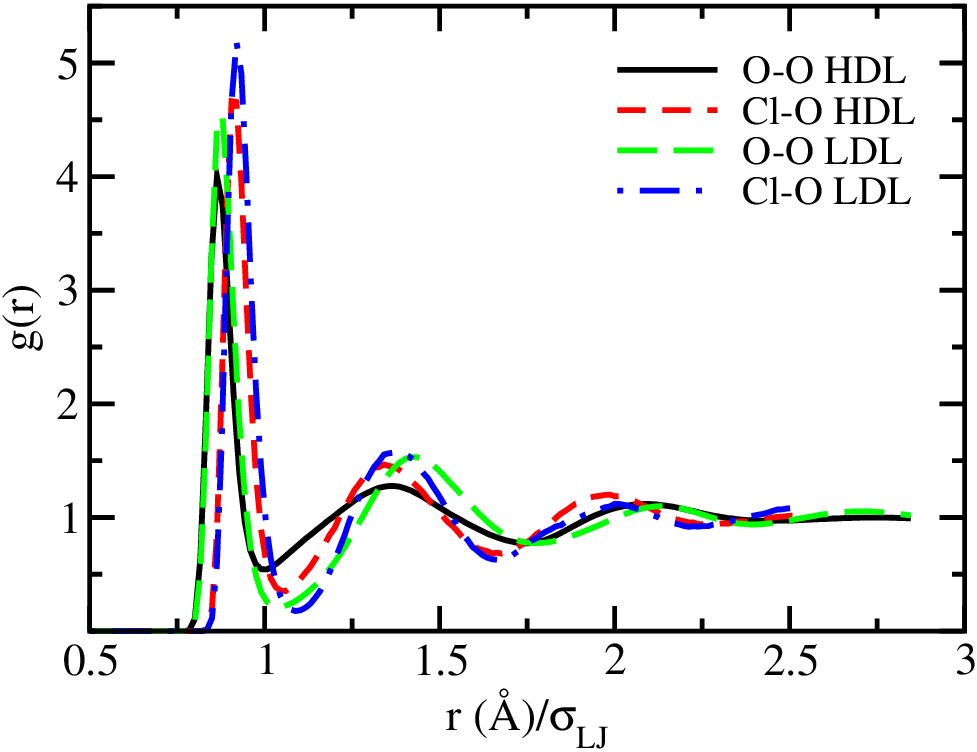}
\caption{Comparison of HDL and LDL O-O and Cl-O RDFs
with distances rescaled by the Lennard-Jones parameter $\sigma$.}
\label{fig:10}
\end{figure}

\begin{table}[htdb]
\caption{First shell, $n_1$,  and second shell, $n_2$, hydration numbers 
for sodium and chloride ions in HDL and LDL at the LLCP
temperature, $T_C =200$~K.}
\begin{center}
\begin{ruledtabular}
\begin{tabular}{lcccc} Atom pair & $n_1$ HDL &$n_1$ LDL & $n_2$ HDL & $n_2$ LDL\\
\hline 
Na-O & 5.975 & 5.957 & 18.490 & 15.898\\
Na-H& 15.912 & 15.364 & 49.474 & 43.521\\
Cl-O& 7.214 & 7.032 & 23.193 & 19.378\\
Cl-H& 6.983 & 7.002 & 33.177 & 27.069\\
\end{tabular}
\end{ruledtabular}
\end{center}
\label{tab:2}
\end{table}

As shown in Fig.~\ref{fig:5} and Fig.~\ref{fig:6}, the ions seem to affect the LDL more 
than the HDL. The most significant difference in the hydration shells of the ions is noticed
in the second shell of the Cl-O RDF. We have also already mentioned that the hydration structure
around the chloride ion resembles the O-O and O-H structures (Fig.~\ref{fig:8}) with the chloride ion able to
substitute oxygen in shells of other oxygen atoms. To further inquire this possibility and to assess
the effect of the chloride ion hydration on HDL/LDL we plot together in Fig.~\ref{fig:10} the O-O
and Cl-O RDFs for HDL and LDL with distances rescaled by the respective LJ interaction distance
parameter $\sigma$. For the O-O pair $\sigma=3.154$~\AA~as given in the TIP4P model~\cite{jorgensen83},
for the Cl-O pair $\sigma=3.561$~\AA~as reported in Table~\ref{tab:1}.

In both LDL and HDL the first and the second shell of oxygen atoms
are closer in the case of the Cl-O RDF 
(see also Fig.~\ref{fig:5} and Fig.~\ref{fig:9}).
To make a quantitative comparison
we can look at the distance between the first peak and the second peak of the RDFs $\Delta R$.
In HDL, for the O-O pair $\Delta R=0.49$, for the Cl-O $\Delta R=0.42$. In LDL,
for the O-O pair $\Delta R=0.55$ and for Cl-O $\Delta R=0.44$ (distances in real units are 
obtained multiplying $\Delta R$ by the respective $\sigma$). Thus we see that in LDL the
chloride ion has the major effect in pulling inward its second hydration shell of oxygens.
As shown in Fig.~\ref{fig:4} and Fig.~\ref{fig:5} the main difference in the structure
of the HDL and LDL is in the position of the second shell of the O-O RDFs, therefore,
a possible explanation for why the presence of ions affect the LDL more than the HDL. 
In fact the chloride ion can take the place of the central oxygen atom in oxygen shells and can
be accommodated in water-water structure at the price of bending hydrogen bonds and pulling
the second shell of oxygen closer to the first shell. This can be tolerated in HDL where the
molecular structure is already collapsed in the second shell~\cite{soper00} but the LDL structure
results instead disrupted. In essence, the substitution of oxygen atoms by chloride atoms together with the hydrogen bond disruption caused by Na$^+$ forces the LDL structure to become more HDL-like. Hence, we can understand also from a structural point
of view why the region of existence of the LDL phase shrinks in the NaCl(aq) with
respect to bulk water.

\section{Conclusions}\label{conclusions}

By the use of MD computer simulations, we have studied the structural properties
of TIP4P bulk water and of a sodium chloride solution in TIP4P water with concentration
$c=0.67$~mol/kg. In particular, we were interested in the structural differences
between the two phases of water, HDL and LDL, that appear in the deep supercooled region.
From our previous work~\cite{corradini10jcp} we knew accurately the phase diagrams
of these systems and in particular the position of the LLCP.

Our results showed that the TIP4P model can reproduce well the structural properties
of the two phases of supercooled liquid water in addition to being a very good potential
for the study of liquid~\cite{corradini10jcp} and solid~\cite{sanz04} water thermodynamics. 
Comparing water-water RDFs in bulk water and in NaCl(aq) we have seen that the LDL
is affected by the presence of ions more than the HDL, as indicated also 
by the shrinkage of the LDL phase observed in the study of the thermodynamics.

The study of the hydration structure of ions in HDL and LDL revealed that a 
disturbance to the LDL structure is induced by the substitution of oxygen by
chloride ions in coordination shells of other oxygen atoms. 
The chloride ions in fact pull inward
its second shell of oxygen atoms, disrupting the LDL structure. 
This, together with the hydrogen bond breaking caused by the sodium ion, 
causes the LDL phase
to be less stable in NaCl(aq) solutions and its region of existence in the 
thermodynamic plane to reduce, with a consequent shift of the liquid-liquid coexistence 
line and of the LLCP to lower pressures with respect to bulk water.

Since from our thermodynamic results we hypothesize that the LLCP region is above
the nucleation line in this solution, an observation of the
structural features presented in this paper in
X-ray and neutron scattering experiments,
can also represent a viable route for experimentalists to solve the quest of
the LLCP in bulk water. Along this line experimental indications of a 
HDL and LDL phase coexistence and LLCP have been 
recently found~\cite{huang10}.

\section*{Acknowledgments}

We gratefully acknowledge the computational resources offered by CINECA
for the ``Progetto Calcolo 891'', by the INFN RM3-GRID at Roma Tre University 
and by the Democritos National Simulation Center at SISSA (Trieste, Italy).

\section*{Appendix}

The TIP4P model for water is a nonpolarizable potential where
three sites are arranged according to the molecular geometry.
The two sites representing the hydrogens are positively charged with 
$q_H=0.52$ $e$; each one forms a rigid bond with the site of the 
oxygen at distance $0.9752$~\AA. The angle between the bonds is $104.52^\circ$.
The site of the oxygen is neutral while a fourth site carries the negative charge of the oxygen
$q_O=-2q_H$. This site is located in the same plane of the molecule at a distance
$0.15$~\AA~from the oxygen with an angle $52.26^\circ$ from the $OH$ bond.
The intermolecular interactions are represented by Eq.~\ref{eq:1}.
The LJ parameters are given by $\sigma_{OO} =3.154$~\AA~and 
$\epsilon_{OO} =0.64852$~kJ/mol.

\end{document}